\PassOptionsToPackage{unicode}{hyperref}
\PassOptionsToPackage{hyphens}{url}
\PassOptionsToPackage{dvipsnames,svgnames,x11names}{xcolor}
\documentclass[
  11pt,
  a4paper,
]{article}
\usepackage{xcolor}
\usepackage[margin=1in]{geometry}
\usepackage{amsmath,amssymb}
\usepackage{iftex}
\ifPDFTeX
  \usepackage[T1]{fontenc}
  \usepackage[utf8]{inputenc}
  \usepackage{textcomp} 
\else 
  \usepackage{unicode-math} 
  \defaultfontfeatures{Scale=MatchLowercase}
  \defaultfontfeatures[\rmfamily]{Ligatures=TeX,Scale=1}
\fi
\usepackage{lmodern}
\ifPDFTeX\else
\fi
\IfFileExists{upquote.sty}{\usepackage{upquote}}{}
\IfFileExists{microtype.sty}{
  \usepackage[]{microtype}
  \UseMicrotypeSet[protrusion]{basicmath} 
}{}
\makeatletter
\@ifundefined{KOMAClassName}{
  \IfFileExists{parskip.sty}{%
    \usepackage{parskip}
  }{
    \setlength{\parindent}{0pt}
    \setlength{\parskip}{6pt plus 2pt minus 1pt}}
}{
  \KOMAoptions{parskip=half}}
\makeatother
\makeatletter
\ifx\paragraph\undefined\else
  \let\oldparagraph\paragraph
  \renewcommand{\paragraph}{
    \@ifstar
      \xxxParagraphStar
      \xxxParagraphNoStar
  }
  \newcommand{\xxxParagraphStar}[1]{\oldparagraph*{#1}\mbox{}}
  \newcommand{\xxxParagraphNoStar}[1]{\oldparagraph{#1}\mbox{}}
\fi
\ifx\subparagraph\undefined\else
  \let\oldsubparagraph\subparagraph
  \renewcommand{\subparagraph}{
    \@ifstar
      \xxxSubParagraphStar
      \xxxSubParagraphNoStar
  }
  \newcommand{\xxxSubParagraphStar}[1]{\oldsubparagraph*{#1}\mbox{}}
  \newcommand{\xxxSubParagraphNoStar}[1]{\oldsubparagraph{#1}\mbox{}}
\fi
\makeatother

\usepackage{longtable,booktabs,array}
\usepackage{calc} 
\usepackage{etoolbox}
\makeatletter
\patchcmd\longtable{\par}{\if@noskipsec\mbox{}\fi\par}{}{}
\makeatother
\IfFileExists{footnotehyper.sty}{\usepackage{footnotehyper}}{\usepackage{footnote}}
\makesavenoteenv{longtable}
\usepackage{graphicx}
\makeatletter
\newsavebox\pandoc@box
\newcommand*\pandocbounded[1]{
  \sbox\pandoc@box{#1}%
  \Gscale@div\@tempa{\textheight}{\dimexpr\ht\pandoc@box+\dp\pandoc@box\relax}%
  \Gscale@div\@tempb{\linewidth}{\wd\pandoc@box}%
  \ifdim\@tempb\p@<\@tempa\p@\let\@tempa\@tempb\fi
  \ifdim\@tempa\p@<\p@\scalebox{\@tempa}{\usebox\pandoc@box}%
  \else\usebox{\pandoc@box}%
  \fi%
}
\def\fps@figure{htbp}
\makeatother

\NewDocumentCommand\citeproctext{}{}

\makeatletter
 \let\@cite@ofmt\@firstofone
 \def\@biblabel#1{}
 \def\@cite#1#2{{#1\if@tempswa , #2\fi}}
\makeatother
\newlength{\cslhangindent}
\newlength{\csllabelwidth}
\newenvironment{CSLReferences}[2] 
 {\begin{list}{}{%
  \setlength{\itemindent}{0pt}
  \setlength{\leftmargin}{0pt}
  \setlength{\parsep}{0pt}
  \ifodd #1
   \setlength{\leftmargin}{\cslhangindent}
   \setlength{\itemindent}{-1\cslhangindent}
  \fi
  \setlength{\itemsep}{#2\baselineskip}}}
 {\end{list}}
\usepackage{calc}

\providecommand{\tightlist}{%
  \setlength{\itemsep}{0pt}\setlength{\parskip}{0pt}}

\makeatletter
\AtBeginDocument{\def\@author{\begin{tabular}[t]{c}Rajius Idzalika, Muhammad Rheza Muztahid, and Radityo Eko Prasojo\\ \small Data Insight for Social and Humanitarian Actions (DISHA)\end{tabular}}}
\makeatother
\makeatletter
\@ifpackageloaded{caption}{}{\usepackage{caption}}
\AtBeginDocument{%
\ifdefined\contentsname
  \renewcommand*\contentsname{Table of contents}
\else
  \newcommand\contentsname{Table of contents}
\fi
\ifdefined\listfigurename
  \renewcommand*\listfigurename{List of Figures}
\else
  \newcommand\listfigurename{List of Figures}
\fi
\ifdefined\listtablename
  \renewcommand*\listtablename{List of Tables}
\else
  \newcommand\listtablename{List of Tables}
\fi
\ifdefined\figurename
  \renewcommand*\figurename{Figure}
\else
  \newcommand\figurename{Figure}
\fi
\ifdefined\tablename
  \renewcommand*\tablename{Table}
\else
  \newcommand\tablename{Table}
\fi
}
\@ifpackageloaded{float}{}{\usepackage{float}}
\floatstyle{ruled}
\@ifundefined{c@chapter}{\newfloat{codelisting}{h}{lop}}{\newfloat{codelisting}{h}{lop}[chapter]}
\floatname{codelisting}{Listing}

\makeatother
\makeatletter
\@ifpackageloaded{caption}{}{\usepackage{caption}}
\@ifpackageloaded{subcaption}{}{\usepackage{subcaption}}
\makeatother
\usepackage{bookmark}
\IfFileExists{xurl.sty}{\usepackage{xurl}}{} 
\hypersetup{
  pdftitle={Context-Aware Displacement Estimation from Mobile Phone Data: A Methodological Framework},
  pdfauthor={Rajius Idzalika; Muhammad Rheza Muztahid; Radityo Eko Prasojo},
  pdfkeywords={displacement, mobile phone data, disaster
response, population mobility, Call Detail Records},
  colorlinks=true,
  linkcolor={blue},
  filecolor={Maroon},
  citecolor={Blue},
  urlcolor={Blue},
  pdfcreator={LaTeX via pandoc}}

\title{Context-Aware Displacement Estimation from Mobile Phone Data: A
Methodological Framework}
\usepackage{etoolbox}
\makeatletter
\providecommand{\subtitle}[1]{
  \apptocmd{\@title}{\par {\large #1 \par}}{}{}
}
\makeatother
\subtitle{Data Insight for Social and Humanitarian Actions (DISHA)}
\author{Rajius Idzalika \and Muhammad Rheza Muztahid \and Radityo Eko
Prasojo}
\date{2026-04-08}
\begin{document}
\maketitle
\begin{abstract}
Timely and accurate population displacement estimates are critical for
humanitarian response during disasters, yet traditional methods such as
surveys and field assessments are resource-intensive and slow. Mobile
phone data offers an alternative for near real-time population movement
tracking. However, existing approaches apply uniform displacement
definitions regardless of individual mobility patterns, leading to
misclassification of regular commuters as displaced persons. We present
a methodological framework that addresses this limitation through three
key innovations: (1) mobility profile classification that distinguishes
local residents from various commuter types, (2) context-aware
between-municipality displacement detection that accounts for expected
location by user type and day of week, and (3) operational uncertainty
bounds derived from baseline coefficient of variation with a disaster
adjustment factor, intended for humanitarian decision support rather
than formal statistical inference. The framework produces three
complementary displacement metrics: displacement rates,
origin-destination flows, and return dynamics, scaled to population
level with uncertainty bounds. The Aparri case study following Super
Typhoon Nando (2025) in the Philippines applies the framework to
vendor-provided daily locations from Globe Telecom, pre-computed from
full 24-hour transaction data. Context-aware detection reduced estimated
between-municipality displacement by 1.6-2.7 percentage points on
weekdays compared to naive methods, a difference attributable to the
commuter exception but not independently validated. The method captures
between-municipality displacement only. Within-municipality evacuation
falls outside its detection scope. The single-case demonstration
establishes proof of concept. External validity requires application
across multiple events and locations. The framework provides
humanitarian actors with operational displacement information while
maintaining individual privacy through aggregation.
\end{abstract}

\section{Introduction}\label{introduction}

\subsection{Problem Statement}\label{problem-statement}

Population displacement during disasters creates urgent humanitarian
needs. Knowing how many people have moved, from where, to where, and
whether they are returning is essential for resource mobilization.
Traditional displacement assessment methods such as household surveys,
registration at displacement sites, and key informant interviews (Expert
Group on Refugee, IDP and Statelessness Statistics, 2023; International
Organization for Migration, 2024; UNHCR, 2023) provide valuable
information but face limitations.

Surveys can take weeks to months to design, deploy, and analyze,
delaying response when speed is critical. Field access is often
restricted in disaster-affected areas, limiting geographic coverage.
Large-scale assessments demand financial and human resources that may
not be available in the immediate aftermath of a disaster. Traditional
methods also produce point-in-time snapshots that miss the evolving
dynamics of displacement, as populations move, shelter temporarily, and
return over days and weeks.

We present a methodological framework for context-aware displacement
estimation from mobile phone data that separates potential displacement
from regular mobility patterns. The framework is designed for future
integration into the DISHA platform, which currently operates on
aggregate mobile data. The methodology extends DISHA toward
individual-level mobility profiling for displacement detection. While we
develop and illustrate the framework using mobile phone data from the
Philippines, where administrative divisions comprise regions (ADM1),
provinces (ADM2), municipalities (ADM3), and barangays (ADM4), the
methodology is designed to be applicable to other contexts with similar
data availability. This paper demonstrates the approach with a single
municipality affected by one typhoon event. Establishing external
validity across multiple events and locations remains future work. We
detail the methodology for mobility profiling, displacement detection,
and uncertainty quantification, then discuss quality assurance
considerations and limitations.

\subsection{Contribution}\label{contribution}

We present a methodological framework that addresses the commuter
misclassification problem through:

\begin{enumerate}
\def\labelenumi{\arabic{enumi}.}
\tightlist
\item
  \textbf{Mobility profile classification}: Categorizing users by their
  baseline weekday vs.~weekend location patterns
\item
  \textbf{Context-aware displacement detection}: Applying different
  displacement rules based on user type and day of week
\item
  \textbf{Empirical uncertainty quantification}: Using coefficient of
  variation from baseline data with disaster adjustment
\end{enumerate}

Validating displacement estimates during emergencies is inherently
difficult: ground truth data is scarce and collecting it diverts
resources from response. Yet the alternative is either waiting weeks for
traditional survey results or operating without displacement
intelligence entirely. The framework provides timely municipality-level
estimates that can inform response while formal validation remains an
ongoing effort.

The framework is designed for sudden-onset disasters with a defined
onset date. Adaptation to other contexts (slow-onset, conflict) would
require methodological modifications discussed in the Generalizability
section.

\section{Background}\label{background}

\subsection{Defining Displacement}\label{defining-displacement}

\subsubsection{Humanitarian Framework}\label{humanitarian-framework}

Displacement is a core concept in humanitarian response. The United
Nations Guiding Principles on Internal Displacement define Internally
Displaced Persons (IDPs) as individuals who have been forced to flee
their homes due to armed conflict, generalized violence, human rights
violations, or natural or human-made disasters, and who have not crossed
an internationally recognized state border (Internal Displacement
Monitoring Centre, 2024; United Nations, 1998). The definition
emphasizes three elements: physical relocation away from habitual
residence (movement), forced or involuntary nature of that movement
(coercion), and remaining within the country of origin (internal).

\subsubsection{Distinguishing Displacement from Other
Mobility}\label{distinguishing-displacement-from-other-mobility}

Not all human movement constitutes displacement.
Table~\ref{tbl-mobility-types} distinguishes displacement from other
forms of population movement.

\begin{longtable}[]{@{}llll@{}}
\caption{Typology of population
movements}\label{tbl-mobility-types}\tabularnewline
\toprule\noalign{}
Movement Type & Voluntary & Temporary & Disaster-Linked \\
\midrule\noalign{}
\endfirsthead
\toprule\noalign{}
Movement Type & Voluntary & Temporary & Disaster-Linked \\
\midrule\noalign{}
\endhead
\bottomrule\noalign{}
\endlastfoot
Daily commuting & Yes & Yes (daily) & No \\
Seasonal migration & Usually & Cyclical & Usually no \\
Permanent relocation & Usually & No & Sometimes \\
Evacuation & Varies & Usually & Yes \\
Disaster displacement & No & Usually & Yes \\
\end{longtable}

\subsubsection{Measurement Challenges}\label{measurement-challenges}

The humanitarian definition of displacement poses measurement
challenges. Intent is unobservable: determining whether movement was
``forced'' or voluntary requires understanding individual motivations,
which cannot be directly measured at scale. Causation is also difficult
to establish, as attributing movement specifically to a disaster event
rather than coincidental travel requires assumptions about temporal
association. Furthermore, continuous monitoring is needed because people
move, shelter temporarily, and return over days and weeks. These
measurement challenges motivate behavioral proxies that infer
displacement from observable location patterns.

\subsection{Mobile Phone Data}\label{mobile-phone-data}

Mobile network operators collect data on subscriber locations through
cell tower connections. The data, when de-identified and aggregated,
offers potential for population movement estimation that addresses
limitations of traditional methods. Prior work has demonstrated the
utility of Call Detail Records (CDR) and similar mobile data for
tracking population movements during disasters, epidemics, and other
crises (Bengtsson et al., 2011; Lu et al., 2012; Wesolowski et al.,
2012).

Mobile phone data offers several advantages for displacement estimation.
Data can be processed much faster than traditional survey-based
approaches, enabling near real-time situational awareness. Because
subscriber connections are recorded continuously, displacement dynamics
can be tracked over time rather than captured at a single point.

Location is recorded at the level of individual cell towers, each of
which falls within an administrative unit, enabling fine spatial
resolution. In our empirical testing with Philippine data, for example,
each tower maps to a barangay (ADM4). However, a tower's coverage area
may be smaller or larger than a single administrative boundary depending
on tower density, coverage range, and antenna directionality (azimuth),
so the effective spatial resolution varies. Even so, mobile data offers
finer resolution than survey-based approaches that usually operate at
ADM2 level. Finally, in countries with high mobile penetration, the data
may capture much of the population.

\subsubsection{Spatial Resolution
Trade-offs}\label{spatial-resolution-trade-offs}

The spatial resolution of mobile phone data depends on cell tower
density and the method used to assign locations to administrative units.
The observed location represents the tower location, not the user's
precise position. Users may be located anywhere within the tower's
coverage area.

In practice, tower coverage areas can be approximated using Voronoi
tessellation, which assigns each point in space to the nearest tower
(Deville et al., 2014). The approximation, however, has known
limitations. Users are often connected to towers outside their nearest
Voronoi polygon due to network load balancing (Ogulenko et al., 2022).
The accuracy of administrative unit assignment then depends on several
interacting factors: tower density, which is typically higher in urban
areas and lower in rural areas; the size of administrative units
relative to tower coverage areas; and the placement of towers relative
to administrative boundaries.

Given the tension between available spatial granularity and the
reliability needed by ground users, we recommend the city or
municipality level (ADM3) as the default spatial unit for displacement
analysis. In the Philippines, ADM3 units are either cities (chartered)
or municipalities. Throughout the methodology we use ``city'' as
shorthand for either type, while the case study uses the legally correct
term for each specific unit (Aparri is a municipality; Tuguegarao is a
city). Barangay level (ADM4) outputs can be produced but should be
validated against tower density maps and interpreted with caution.

\subsubsection{Displacement Proxy}\label{displacement-proxy}

Mobile phone data enables a behavioral proxy approach to displacement
measurement. Since intent and causation cannot be directly observed, we
operationalize displacement as anomalous absence from expected location:

\begin{quote}
\textbf{Operational definition}: A user is classified as displaced on
day \(t\) if their observed location differs from their expected
location given their established mobility profile and the day of week,
during the post-disaster observation period.
\end{quote}

The definition has three key components. The established mobility
profile is derived from the pre-disaster baseline period and captures
each user's regular location patterns. The expected location is then
determined by the user's type (local resident, commuter, etc.) and the
day of week, accounting for predictable weekday-weekend differences.
Finally, displacement classification only applies during the
post-disaster observation period, establishing temporal association with
the disaster event.

\subsubsection{Displacement Proxy
Limitations}\label{displacement-proxy-limitations}

Using location anomaly as a proxy for displacement introduces several
limitations. First, the approach cannot confirm involuntary movement.
For example, a person who leaves their home city after a disaster may be
displaced (whether through organized evacuation or spontaneous flight)
or visiting relatives for unrelated reasons. Second, the framework
assumes location anomaly is an indicator of displacement. Nevertheless,
causation (that the disaster causes people to move) cannot be proven
from observational data alone.

The estimates represent potential displacement: the population whose
location behavior is consistent with displacement, not confirmed
displaced persons. The approach is analogous to excess mortality
estimation, where we measure deviation from expected patterns rather
than verified individual cases. Additionally, the approach excludes
immobile affected populations, such as persons who remain in
disaster-affected areas. These persons are not captured as displaced,
though they may have humanitarian needs.

Despite the limitations, behavioral proxies provide situational
awareness when direct measurement is infeasible, particularly in the
immediate post-disaster period when humanitarian decisions must be made
rapidly.

\subsection{Gap in Current Approaches}\label{gap-in-current-approaches}

Existing mobile phone-based displacement methods typically define
displacement using a uniform rule: a person is displaced if their
observed location differs from their home location. The definition does
not account for regular commuting patterns.

Consider an inter-city commuter who lives in City A but works in City B.
Under the uniform rule:

\begin{itemize}
\tightlist
\item
  \textbf{Every weekday}, they are observed in City B (their workplace),
  which differs from their home in City A
\item
  The method flags them as ``displaced'' on normal workdays---a
  \textbf{false positive}
\end{itemize}

The uniform approach leads to systematic overestimation:

\begin{itemize}
\tightlist
\item
  \textbf{False positives}: Commuters flagged as displaced during
  routine work travel
\item
  \textbf{Biased estimates}: Errors concentrate in populations with high
  commuting rates
\end{itemize}

A separate limitation affects all location-based approaches, including
ours. If a displaced person coincidentally ends up at their usual
expected location (e.g., a commuter displaced to their normal work
city), they would not be counted as displaced. The false negative
problem cannot be resolved without additional information about
displacement intent.

\section{Related Work}\label{related-work}

\subsection{Mobile Phone Data for Population
Mobility}\label{mobile-phone-data-for-population-mobility}

The use of mobile phone data for population mobility analysis has grown
since the early 2010s. Gonzalez et al. (2008) demonstrated that
individual human mobility patterns are highly predictable, with most
people showing a high probability of returning to a small number of
frequently visited locations.

\subsection{Displacement Estimation from
CDR}\label{displacement-estimation-from-cdr}

Bengtsson et al. (2011) pioneered the use of CDR for disaster
displacement estimation following the 2010 Haiti earthquake,
demonstrating that mobile phone data could provide population movement
estimates within days of a disaster. Lu et al. (2012) extended the work
to examine the predictability of displacement patterns, finding that
pre-disaster mobility patterns predicted post-disaster destinations.
More recently, Khaefi et al. (2018) applied machine learning to mobile
network data, demonstrating that spatial behavior features could predict
evacuation destinations before disasters occur.

The Flowminder Foundation has developed operational methods for
CDR-based population estimation, applying the techniques across multiple
disaster contexts including the 2015 Nepal earthquake (R. Wilson et al.,
2016). Their approaches define home location as the modal nighttime
location over a baseline period. Flowminder's methodology represents a
widely-used operational approach: displacement is detected when a user's
observed location differs from their assigned home location. The
approach is effective for populations with limited regular mobility but
does not distinguish between displacement and routine commuting. Our
framework builds on Flowminder's home detection methods while adding
mobility profile classification to reduce false positives among commuter
populations.

\subsection{General Population Mobility
Frameworks}\label{general-population-mobility-frameworks}

Ahas et al. (2010) and colleagues at Positium developed methods for
passive mobile positioning that model locations meaningful to users of
mobile phones. Their work emphasizes the importance of temporal patterns
(time of day, day of week) in interpreting location data.

\subsection{Gap Addressed}\label{gap-addressed}

While prior work establishes robust methods for home location detection
and aggregate mobility estimation, existing approaches do not account
for commuting patterns when defining displacement. Our framework extends
the methodological tradition developed by Flowminder and codified by
Arai et al. (2022), introducing explicit user classification and
context-aware displacement logic.

\section{Data and Privacy}\label{data-and-privacy}

\subsection{Data Description}\label{data-description}

The methodology requires mobile phone transaction data with the
following minimum fields:

\begin{itemize}
\tightlist
\item
  \textbf{User identifier}: Pseudonymized subscriber ID (cryptographic
  hash)
\item
  \textbf{Timestamp}: Date and time of network activity
\item
  \textbf{Location}: Cell tower or administrative unit identifier
\end{itemize}

Additional fields (call type, duration, etc.) are not required.

\subsection{Privacy Safeguards}\label{privacy-safeguards}

The methodology requires persistent user identifiers to link
observations over time: establishing baseline home locations, tracking
post-disaster displacement, and measuring return dynamics all depend on
following the same individual across days and weeks. Without persistent
identifiers, the framework cannot function.

However, because persistent identifiers remain linkable over time, the
data are pseudonymized rather than anonymized. The framework
incorporates privacy protection through multiple safeguards:

\begin{itemize}
\tightlist
\item
  \textbf{Pseudonymization}: User identifiers are cryptographically
  hashed before data reaches the analysis team. Researchers never access
  direct identifiers (phone numbers). The data provider (mobile network
  operator) maintains exclusive control of the re-identification key.
\item
  \textbf{Aggregation}: All published outputs are aggregated to
  administrative unit level (city/municipality minimum). No individual
  trajectories appear in outputs.
\item
  \textbf{Minimum threshold suppression}: Results are suppressed when
  cell counts fall below minimum thresholds. Research has shown that
  human mobility traces are highly unique: four spatio-temporal points
  can uniquely identify 95\% of individuals (Montjoye et al., 2013),
  which necessitates careful aggregation.
\item
  \textbf{Temporal resolution}: Daily aggregates reduce
  re-identification risk compared to finer-grained timestamps.
\item
  \textbf{Data governance}: Analysis occurs within secure environments.
  Only aggregated metrics are exported.
\end{itemize}

\section{Methodology}\label{methodology}

\subsection{Overview}\label{overview}

The methodology proceeds through the following stages:

\begin{enumerate}
\def\labelenumi{\arabic{enumi}.}
\tightlist
\item
  \textbf{Daily location signal}: Obtain one location per user per day.
  The framework's external mode (used in this paper) consumes a
  vendor-supplied daily aggregation. An internal mode that would compute
  separate residential and daytime-activity signals from intra-day
  inputs is described as a design specification in §5.4 but is not
  exercised in this paper.
\item
  \textbf{Residential baseline establishment}: Determine stable ``true
  home'' from baseline period
\item
  \textbf{Mobility profile classification}: Categorize users by
  commuting pattern
\item
  \textbf{Displacement detection}: Apply context-aware rules based on
  user type and day
\item
  \textbf{Metric calculation}: Compute displacement rates, O-D flows,
  and return dynamics
\item
  \textbf{Population scaling}: Scale mobile counts to population with
  uncertainty bounds
\end{enumerate}

\begin{longtable}[]{@{}
  >{\raggedright\arraybackslash}p{(\linewidth - 4\tabcolsep) * \real{0.2692}}
  >{\raggedright\arraybackslash}p{(\linewidth - 4\tabcolsep) * \real{0.2308}}
  >{\raggedright\arraybackslash}p{(\linewidth - 4\tabcolsep) * \real{0.5000}}@{}}
\caption{Methodology pipeline from raw mobile data to population-scaled
displacement metrics}\label{tbl-pipeline}\tabularnewline
\toprule\noalign{}
\begin{minipage}[b]{\linewidth}\raggedright
Stage
\end{minipage} & \begin{minipage}[b]{\linewidth}\raggedright
Step
\end{minipage} & \begin{minipage}[b]{\linewidth}\raggedright
Description
\end{minipage} \\
\midrule\noalign{}
\endfirsthead
\toprule\noalign{}
\begin{minipage}[b]{\linewidth}\raggedright
Stage
\end{minipage} & \begin{minipage}[b]{\linewidth}\raggedright
Step
\end{minipage} & \begin{minipage}[b]{\linewidth}\raggedright
Description
\end{minipage} \\
\midrule\noalign{}
\endhead
\bottomrule\noalign{}
\endlastfoot
\textbf{1. Input} & Data preparation & One vendor-supplied location per
user per day (external mode) \\
\textbf{2. Baseline Establishment} & 2.1 Residential baseline &
Aggregate to stable home (weekend priority, weekday fallback) \\
& 2.2 Mobility profile & Classify users by commuting pattern \\
\textbf{3. Displacement Detection} & 3.1 Context-aware rules & Apply
displacement logic by user type and day of week \\
& 3.2 Metric calculation & Compute rates, O-D flows, return dynamics \\
\textbf{4. Output} & 4.1 Population scaling & Scale mobile counts using
population/subscriber ratio \\
& 4.2 Uncertainty bounds & Apply CV-based uncertainty bounds \\
\end{longtable}

\begin{figure}

\centering{

\includegraphics[width=6in,height=3.17in]{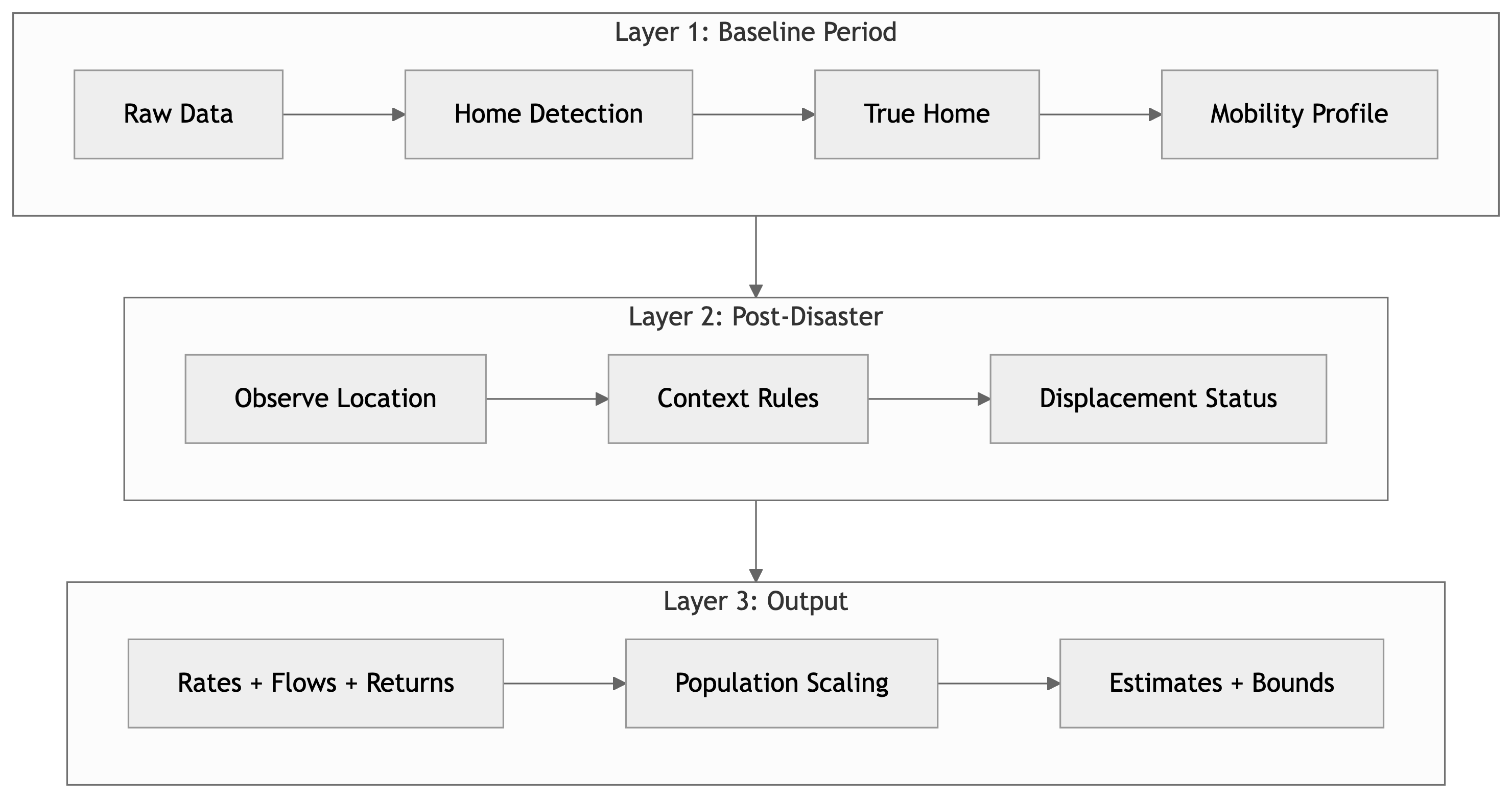}

}

\caption{\label{fig-pipeline}Context-aware displacement estimation
pipeline.}

\end{figure}%

\subsection{Temporal Framework}\label{temporal-framework}

The methodology requires definition of three time periods:

\begin{itemize}
\tightlist
\item
  \textbf{Baseline period}: Pre-disaster period for establishing home
  locations and mobility patterns. We recommend a minimum of 6 weeks for
  home detection and mobility profiling, following Bengtsson et al.
  (2011) who used 42 days in the Haiti earthquake response. This
  duration captures recurring weekly patterns while remaining short
  enough for rapid processing. A longer separate baseline (ideally 3
  months or more) is preferable for estimating the city-level
  coefficient of variation used in the uncertainty bounds of §5.7, since
  CV estimates from short windows can be unstable. Days coinciding with
  national holidays should be excluded from both baselines, as they
  reflect anomalous rather than typical mobility.
\item
  \textbf{Disaster onset}: Date/time when displacement may begin
\item
  \textbf{Post-disaster observation}: Period for tracking displacement
  and returns
\end{itemize}

\subsection{Daily Location Signals}\label{daily-location-signals}

Commuter-aware displacement detection requires distinguishing where a
user \emph{lives} from where they are \emph{during the day}. The
framework supports two modes. In \textbf{external detection} (used in
this paper), it receives a single vendor-supplied location per day and
treats weekend values as residential-like (people are typically home on
weekends) and weekday values as daytime-activity-like. In
\textbf{internal detection} (design only, not exercised here), it
computes separate residential and daytime-activity signals from
intra-day inputs (event-level CDRs, hourly aggregates, or similar),
using a nighttime-priority procedure for the residential signal and a
daytime modal for the activity signal. Tie-breaking uses a fixed random
seed. Each residential assignment would carry a quality flag
(Table~\ref{tbl-quality-classification}). Various existing
home-detection algorithms could fill this role (Nugroho et al., 2021;
Vanhoof et al., 2018).

Whether external-mode commuter classification succeeds depends on
whether the vendor signal shifts with day-of-week. Daily commuters (who
sleep at home every night) produce day-of-week variation only if the
vendor aggregation captures daytime activity: mobile transactions
concentrate during working hours, so a modal-over-transactions
aggregation can register the work city more strongly on weekdays than a
time-weighted average would. Weekly commuters (who stay at the work city
Mon-Fri including sleeping there) produce day-of-week variation under
any scheme, because their sleeping location itself moves between cities.
The sensitivity is therefore asymmetric: a purely nighttime-weighted
vendor scheme would detect weekly commuters but miss daily commuters
entirely. Internal detection would avoid this asymmetry by design, but
the comparison is theoretical. Practitioners should assess the condition
by comparing each user's modal weekday and weekend values. If no users
show variation, daily commuters are not being detected. If some users do
show variation (§6.2), the variation confirms the vendor signal is not
degenerate, but it may reflect weekly commuters alone rather than daily
commuters.

\begin{longtable}[]{@{}lll@{}}
\caption{Quality classification for internal-detection residential
signals. In external-detection mode, the flag is not available because
the aggregation happens inside the vendor
pipeline.}\label{tbl-quality-classification}\tabularnewline
\toprule\noalign{}
Quality Level & Definition & Interpretation \\
\midrule\noalign{}
\endfirsthead
\toprule\noalign{}
Quality Level & Definition & Interpretation \\
\midrule\noalign{}
\endhead
\bottomrule\noalign{}
\endlastfoot
High & Nighttime location data available & Strong residential signal \\
Medium & Only daytime data available & Moderate confidence \\
Low & Insufficient data in either period & Weak signal \\
\end{longtable}

\subsection{Residential Baseline
Establishment}\label{sec-residential-baseline}

Weekday locations are confounded by work. A person's most frequent
weekday location may be their workplace rather than residence. Weekend
locations more reliably reflect residential location, as most people
return home on weekends regardless of commute patterns. We therefore
prioritize weekend observations for establishing the residential
baseline.

For each user, we collect the daily residential signals from weekend
days (Saturday, Sunday) in the baseline period and identify the most
frequent value. We require ≥2 weekend days with data for a valid weekend
baseline. The threshold filters out weekend travelers based on
probabilistic reasoning: it is unlikely that a visitor would (a) appear
at the same non-home location on two weekend days, (b) have their signal
recorded there both times, and (c) have no signal at their actual home
during those periods. The combination of conditions is improbable for
non-residents, making ≥2 weekend observations a reasonable filter.

For users without sufficient weekend data, we fall back to the modal
weekday vendor value with ≥5 days minimum, used as a fallback
residential proxy under the same caveat as in §5.4 (we cannot verify the
vendor weighting). Similar probabilistic reasoning applies. The ≥5
threshold is equivalent to one work week, providing enough observations
to establish a recurring pattern while filtering out occasional
travelers.

\begin{equation}\phantomsection\label{eq-baseline}{
\text{Residential Baseline} =
\begin{cases}
\text{Mode}(\text{Weekend Locations}) & \text{if weekend days} \geq 2 \\
\text{Mode}(\text{Weekday Locations}) & \text{if weekend days} < 2 \text{ and weekday days} \geq 5 \\
\text{Excluded} & \text{otherwise}
\end{cases}
}\end{equation}

Each user's residential baseline records its derivation source (weekend
or weekday fallback) for quality assessment and sensitivity analysis.

\subsection{Mobility Profile
Classification}\label{mobility-profile-classification}

Research on human mobility has identified distinct behavioral classes:
``returners'' who exhibit highly predictable patterns centered on a few
locations, and ``explorers'' with more varied movement (Pappalardo et
al., 2015). We classify users into mobility profiles by comparing the
residential baseline (weekend home) with the modal weekday value of the
daytime activity signal. When multiple weekday values have equal
frequency, we select one uniformly at random with fixed seed. The
analysis operates at city level, so commuters whose weekday value is in
a different city are all treated as inter-city commuters, regardless of
whether they cross provincial or regional boundaries.

\begin{longtable}[]{@{}
  >{\raggedright\arraybackslash}p{(\linewidth - 4\tabcolsep) * \real{0.3673}}
  >{\raggedright\arraybackslash}p{(\linewidth - 4\tabcolsep) * \real{0.2449}}
  >{\raggedright\arraybackslash}p{(\linewidth - 4\tabcolsep) * \real{0.3878}}@{}}
\caption{Mobility profile classification based on residential baseline
and weekday location.}\label{tbl-mobility-profiles}\tabularnewline
\toprule\noalign{}
\begin{minipage}[b]{\linewidth}\raggedright
Mobility Profile
\end{minipage} & \begin{minipage}[b]{\linewidth}\raggedright
Definition
\end{minipage} & \begin{minipage}[b]{\linewidth}\raggedright
Expected Behavior
\end{minipage} \\
\midrule\noalign{}
\endfirsthead
\toprule\noalign{}
\begin{minipage}[b]{\linewidth}\raggedright
Mobility Profile
\end{minipage} & \begin{minipage}[b]{\linewidth}\raggedright
Definition
\end{minipage} & \begin{minipage}[b]{\linewidth}\raggedright
Expected Behavior
\end{minipage} \\
\midrule\noalign{}
\endhead
\bottomrule\noalign{}
\endlastfoot
Local Resident & Same barangay on weekdays and weekends & Present in
home area daily \\
Intra-City Commuter & Different barangay, same city & Moves within city
for work \\
Inter-City Commuter & Different city (any distance) & Commutes to
another city for work \\
Weekend-Only User & No weekday data available & Only observed on
weekends \\
\end{longtable}

\subsection{Displacement Detection}\label{displacement-detection}

The key methodological innovation is applying different displacement
rules based on user type and day of week.

\begin{longtable}[]{@{}
  >{\raggedright\arraybackslash}p{(\linewidth - 6\tabcolsep) * \real{0.2727}}
  >{\raggedright\arraybackslash}p{(\linewidth - 6\tabcolsep) * \real{0.1515}}
  >{\raggedright\arraybackslash}p{(\linewidth - 6\tabcolsep) * \real{0.3182}}
  >{\raggedright\arraybackslash}p{(\linewidth - 6\tabcolsep) * \real{0.2576}}@{}}
\caption{Displacement detection logic by mobility profile and day
type}\label{tbl-displacement-matrix}\tabularnewline
\toprule\noalign{}
\begin{minipage}[b]{\linewidth}\raggedright
Mobility Profile
\end{minipage} & \begin{minipage}[b]{\linewidth}\raggedright
Day Type
\end{minipage} & \begin{minipage}[b]{\linewidth}\raggedright
Expected Location(s)
\end{minipage} & \begin{minipage}[b]{\linewidth}\raggedright
Displaced If\ldots{}
\end{minipage} \\
\midrule\noalign{}
\endfirsthead
\toprule\noalign{}
\begin{minipage}[b]{\linewidth}\raggedright
Mobility Profile
\end{minipage} & \begin{minipage}[b]{\linewidth}\raggedright
Day Type
\end{minipage} & \begin{minipage}[b]{\linewidth}\raggedright
Expected Location(s)
\end{minipage} & \begin{minipage}[b]{\linewidth}\raggedright
Displaced If\ldots{}
\end{minipage} \\
\midrule\noalign{}
\endhead
\bottomrule\noalign{}
\endlastfoot
Local Resident & Any & Home city & Outside home city \\
Intra-City Commuter & Any & Home city & Outside home city \\
Inter-City Commuter & Weekend & Home city & Outside home city \\
Inter-City Commuter & Weekday & Home city OR work city & Outside both
cities \\
Weekend-Only User & Any & Home city & Outside home city \\
\end{longtable}

Users present in the baseline but absent from post-disaster observations
are classified as ``unobservable.'' Unobservability may indicate no
active phone transactions during the observation period, movement
outside the geographic coverage area (such as rural areas without tower
coverage or overseas travel), or damage to telecommunications
infrastructure in their area. We report the count and proportion of
unobservable users separately.

\subsection{Displacement Metrics}\label{displacement-metrics}

The framework produces three complementary metrics. \textbf{Displacement
rates} (Equation~\ref{eq-displacement-rate}) measure the proportion of a
city's baseline population that is displaced on each day.

For each city \(c\) and post-disaster day \(t\), we define:

\begin{itemize}
\tightlist
\item
  \(N_c\) = count of users with residential baseline in city \(c\)
  (baseline users)
\item
  \(M_{c,t}\) = count of baseline users not observed in any location on
  day \(t\) (missing users)
\item
  \(D_{c,t}\) = count of baseline users observed on day \(t\) at a
  location other than their expected location, per
  Table~\ref{tbl-displacement-matrix} (displaced users)
\end{itemize}

The displacement rate and missing rate for origin city \(c\) on day
\(t\) are:

\begin{equation}\phantomsection\label{eq-displacement-rate}{
\text{Displacement Rate}_{c,t} = \frac{D_{c,t}}{N_c} \times 100\%
}\end{equation}

\begin{equation}\phantomsection\label{eq-missing-rate}{
\text{Missing Rate}_{c,t} = \frac{M_{c,t}}{N_c} \times 100\%
}\end{equation}

The displacement rate is a potential-displacement estimate in the proxy
sense of §2.3. Two distinct sources of uncertainty affect the estimate,
and the framework treats them separately.

\textbf{Partial identification due to unobserved users.} Missing users
cannot be classified as displaced or not displaced since their location
is unknown. The potential displacement rate lies between the observed
rate (lower bound, if no missing user is displaced) and the observed
rate plus the missing rate (upper bound, if every missing user is
displaced). The interval is a partial identification region, not a
confidence interval: its width depends on data availability, not
sampling variability.

\textbf{Conditional variability among observed users.} Daily counts
fluctuate even in normal periods. Given a longer historical baseline (3
months or more), practitioners can compute the city-level coefficient of
variation (CV) from daily user counts and attach operational bounds to
displacement rates (Equation~\ref{eq-cv}, Equation~\ref{eq-rate-ci}). CV
is the ratio of standard deviation to mean. To account for
crisis-induced behavioral variability, baseline CV is multiplied by a
disaster adjustment factor (default 2):

\begin{equation}\phantomsection\label{eq-cv}{
\text{CV}_{\text{disaster}} = \text{CV}_{\text{baseline}} \times \text{Disaster Multiplier}
}\end{equation}

\begin{equation}\phantomsection\label{eq-rate-ci}{
\text{Uncertainty Bounds} = \text{Displacement Rate}_{c,t} \times (1 \pm 1.96 \times \text{CV}_{\text{disaster}})
}\end{equation}

The multiplier 1.96 derives from the normal distribution (the z-score
for 95\% coverage) but is used here as a conventional scaling factor
rather than a formal statistical claim. Practitioners may adjust based
on decision context: 1.645 for narrower bounds when speed is critical,
or 2.576 for more conservative estimates when resource misallocation is
costly.

Binomial intervals such as Clopper-Pearson (Clopper \& Pearson, 1934)
and Wilson (E. B. Wilson, 1927) quantify within-day sampling precision
rather than the baseline day-to-day variability that matters
operationally. With large \(n\), the binomial standard error becomes
small regardless of the actual day-to-day fluctuation. Overdispersion
adjustments such as quasi-likelihood (Wedderburn, 1974) and
design-effect (Kish, 1965) corrections address within-day clustering but
remain anchored to within-day sampling variance. We therefore use
CV-based bounds as a decision-support measure, anchoring interval width
to baseline day-to-day variability. Monte Carlo bootstrap can be used as
a cross-check, but it is not practical for real-time operations.

The CV-based approach is best suited to medium-rate regimes (roughly 2\%
to 30\%). At very low rates, the multiplicative interval width collapses
toward zero; at very high rates, the upper bound approaches the 100\%
ceiling. A Beta-based formulation would address both boundary cases and
is left for future work. When baselines are short (e.g., 6 weeks), CV
estimates may be unreliable, and practitioners should report the
displacement rate together with the missing rate rather than relying on
fragile uncertainty bounds. The missing rate should be interpreted
separately, as it reflects partial identification from unobserved users
rather than conditional variability among observed users.

\textbf{Origin-destination flows} (Equation~\ref{eq-flow}) capture where
displaced persons go. The flow restricts to users classified as
displaced on day \(t\) by the context-aware rules. Commuters at their
expected weekday locations are excluded. For each origin-destination
city pair \((o, d)\) on day \(t\):

\begin{equation}\phantomsection\label{eq-flow}{
\text{Flow}_{o,d,t} = \text{Count of users with residential baseline } o, \text{ classified as displaced, observed in } d \text{ on day } t
}\end{equation}

Flow counts are aggregated into an O-D matrix showing displacement
patterns between cities. Missing users are excluded since their
destination cannot be observed. When the missing rate is high, observed
flows likely underestimate actual displacement movement. Practitioners
should interpret flow magnitudes in context of the origin city's missing
rate. High missing rates suggest the true flows may be substantially
larger than reported.

\textbf{Return dynamics} (Equation~\ref{eq-returned},
Equation~\ref{eq-cumulative-return}) track how displaced populations
return home over time. For each origin city \(c\) on day \(t\):

\begin{equation}\phantomsection\label{eq-returned}{
R_{c,t} = \text{Count of users displaced on } t-1 \text{ and observed at expected location on } t
}\end{equation}

\begin{equation}\phantomsection\label{eq-max-displaced}{
\text{Max Displaced}_c = \max_{t} \left( D_{c,t} + \sum_{s=1}^{t} R_{c,s} \right)
}\end{equation}

\(\text{Max Displaced}_c\) aggregates currently displaced users with
cumulative return events through day \(t\), then takes the peak across
days. Under monotone classification (each user transitions displaced →
returned at most once), the quantity equals the number of unique users
ever displaced. When users can be re-displaced or classification can
fluctuate, the same user contributes multiple return events, and the
quantity becomes a person-event measure of cumulative displacement
intensity rather than a count of unique individuals. We use the formula
as-is for simplicity and report the resulting return rate as a recovery
indicator, not a strict unique-user fraction.

\begin{equation}\phantomsection\label{eq-cumulative-return}{
\text{Cumulative Return Rate}_{c,t} = \frac{\sum_{s=1}^{t} R_{c,s}}{\text{Max Displaced}_c} \times 100\%
}\end{equation}

Because \(\text{Max Displaced}_c\) takes a maximum over the full
observation period, Equation~\ref{eq-cumulative-return} is a
\emph{retrospective} indicator: the denominator is fixed only after the
period ends, and the series is computed post hoc. The formula is
therefore suited to after-action recovery analysis, not to real-time
monitoring during a disaster. An operational variant for real-time use
would replace the denominator with a running maximum
\(\text{Max Displaced}_{c, t} = \max_{s \leq t}(D_{c,s} + \sum_{u \leq s} R_{c,u})\),
which uses only information available up to day \(t\) but produces a
non-monotonic series that is harder to interpret. We do not pursue the
operational variant in this paper and flag it as a possible extension.

Handling of missing users is asymmetric. The numerator requires
confirmed observation at the expected location, while the denominator
includes all displacement events (including users now missing). The
asymmetry produces a conservative return rate estimate.

\subsection{Population Scaling with
Uncertainty}\label{population-scaling-with-uncertainty}

This section produces an \textbf{estimated displaced population count}:
the number of people-equivalents displaced from city \(c\) on day \(t\),
distinct from the subscriber-level rate in §5.7. Mobile phone
subscribers represent a sample, not a census, so we apply a scaling
factor.

The \textbf{scaling factor} (Equation~\ref{eq-scaling}) for each city is
the ratio of population to average daily baseline subscribers:

\begin{equation}\phantomsection\label{eq-scaling}{
\text{Scaling Factor}_c = \frac{\text{Population}_c}{\text{Avg Daily Baseline}_c}
}\end{equation}

where Average Daily Baseline =
\(\frac{(\text{Weekday Baseline} \times 5) + (\text{Weekend Baseline} \times 2)}{7}\),
accounting for the different activity patterns between weekdays and
weekends. The denominator is the average daily count of active baseline
subscribers, not the unique cohort size \(N_c\), because the scaling
factor maps subscribers typically observed on a day to the city's
resident population.

The simple scaling approach has trade-offs. The method is transparent,
requires minimal data, and is computationally efficient. However, the
approach assumes uniform mobile penetration within each city and does
not account for demographic biases in phone ownership. More
sophisticated approaches such as demographic stratification or
propensity weighting could address the limitations but require
additional data that may not be available in emergency contexts. The
simple scaling factor provides a first-order approximation suitable for
rapid humanitarian response.

\(\text{CV}_{\text{disaster}}\) (§5.7) also applies to population-scaled
estimates. For a subscriber displacement count \(\hat{D}\):

\begin{equation}\phantomsection\label{eq-pop-estimate}{
\text{Estimated Displaced Population} = \hat{D} \times \text{Scaling Factor}_c
}\end{equation}

\begin{equation}\phantomsection\label{eq-pop-ci}{
\text{Estimated Displaced Population Bounds} = \hat{D} \times (1 \pm 1.96 \times \text{CV}_{\text{disaster}}) \times \text{Scaling Factor}_c
}\end{equation}

The scaling factor multiplies both the estimate and its bounds linearly,
so Equation~\ref{eq-pop-ci} inherits the scope and limitations of §5.7.
If a population-level rate is desired, dividing by city population gives
\(\hat{D} / \text{Avg Daily Baseline}_c\), which uses a slightly
different denominator than the subscriber rate in §5.7.

\subsection{Parameter Summary}\label{parameter-summary}

The methodology involves several configurable parameters with default
values chosen based on domain knowledge and practical considerations.
Practitioners should consider adjusting them based on local context. For
example, using different nighttime windows for populations with high
night-shift prevalence, or adjusting observation thresholds based on
data density.

\begin{longtable}[]{@{}
  >{\raggedright\arraybackslash}p{(\linewidth - 6\tabcolsep) * \real{0.2391}}
  >{\raggedright\arraybackslash}p{(\linewidth - 6\tabcolsep) * \real{0.3261}}
  >{\raggedright\arraybackslash}p{(\linewidth - 6\tabcolsep) * \real{0.2391}}
  >{\raggedright\arraybackslash}p{(\linewidth - 6\tabcolsep) * \real{0.1957}}@{}}
\caption{Key parameters and default
values}\label{tbl-parameters}\tabularnewline
\toprule\noalign{}
\begin{minipage}[b]{\linewidth}\raggedright
Parameter
\end{minipage} & \begin{minipage}[b]{\linewidth}\raggedright
Default Value
\end{minipage} & \begin{minipage}[b]{\linewidth}\raggedright
Rationale
\end{minipage} & \begin{minipage}[b]{\linewidth}\raggedright
Tunable
\end{minipage} \\
\midrule\noalign{}
\endfirsthead
\toprule\noalign{}
\begin{minipage}[b]{\linewidth}\raggedright
Parameter
\end{minipage} & \begin{minipage}[b]{\linewidth}\raggedright
Default Value
\end{minipage} & \begin{minipage}[b]{\linewidth}\raggedright
Rationale
\end{minipage} & \begin{minipage}[b]{\linewidth}\raggedright
Tunable
\end{minipage} \\
\midrule\noalign{}
\endhead
\bottomrule\noalign{}
\endlastfoot
Nighttime window & 21:00-04:59 & Aligns with typical sleep hours &
Yes \\
Weekend threshold & ≥2 days & Minimum for mode stability & Yes \\
Weekday threshold & ≥5 days & Compensates for commuting noise & Yes \\
Disaster CV multiplier & 2.0 & Accounts for crisis-induced variability &
Yes \\
\end{longtable}

\section{Case Study: Super Typhoon Nando (Aparri,
2025)}\label{case-study-super-typhoon-nando-aparri-2025}

To demonstrate the framework, we apply it to displacement estimation for
Aparri, Cagayan during Super Typhoon Nando (international name: Ragasa)
in September 2025.

\subsection{Event Description}\label{event-description}

Super Typhoon Nando made landfall over Calayan, Cagayan on September 22,
2025 at approximately 5:00 PM local time with sustained winds of 215
km/h (Philippine Atmospheric, Geophysical and Astronomical Services
Administration, 2025). The typhoon caused damage across the Cagayan
Valley region, particularly in coastal municipalities including Aparri.

\subsection{Data and Baseline
Characteristics}\label{data-and-baseline-characteristics}

We analyzed mobile phone data from August 10 to October 6, 2025, for all
subscribers who visited Aparri at least once during the observation
period. Globe Telecom supplied one location per user per day,
pre-computed from the full day's transaction records.

Because Globe's weighting scheme is proprietary, the weekday-weekend
variation observed for 12.1\% of users
(Table~\ref{tbl-user-distribution}) is consistent with at least weekly
commuting and possibly daily commuting (§5.4), but does not identify the
weighting rule itself. Alternative explanations (non-commuting weekly
routines, irregular work schedules, regular weekend travel) cannot be
ruled out without individual-level ground truth.

The baseline period (August 10 - September 21, 40 days after excluding
three national holidays) established home locations and mobility
profiles for tens of thousands of subscribers detected in 98
municipalities. Exact subscriber counts are withheld to protect
proprietary data. All results are reported as percentages. The
post-disaster observation period covers September 22 - October 6 (15
days). Population-level estimates scale the sample using
municipality-level subscriber-to-population ratios. Population counts
come from WorldPop 2024 (WorldPop, 2024) spatially joined to the 2023
Philippine administrative boundary map (National Mapping and Resource
Information Authority, 2023). PSA Census 2020 (Philippine Statistics
Authority, 2021) serves as fallback for administrative units that cannot
be matched in the spatial join.

Missing subscribers (those unobservable on a given post-disaster day)
may reflect infrastructure damage, power outages, or evacuation to areas
without network coverage. The missing rate is reported alongside
displacement rates in Table~\ref{tbl-naive-comparison} and used to
construct scenario bounds in Section 6.4.

\begin{longtable}[]{@{}ll@{}}
\caption{Pipeline attrition for the Aparri case study, as percentages of
the starting population (subscribers ever observed in Aparri during the
baseline period).}\label{tbl-attrition}\tabularnewline
\toprule\noalign{}
Pipeline stage & \% of starting population \\
\midrule\noalign{}
\endfirsthead
\toprule\noalign{}
Pipeline stage & \% of starting population \\
\midrule\noalign{}
\endhead
\bottomrule\noalign{}
\endlastfoot
Subscribers observed in Aparri during baseline & 100.0\% \\
Valid residential baseline (≥2 weekend or ≥5 weekday days) & 96.2\% \\
Observed at least once during post-disaster period & 90.8\% \\
\end{longtable}

\begin{longtable}[]{@{}ll@{}}
\caption{Distribution of subscribers by mobility profile in the Aparri
baseline sample}\label{tbl-user-distribution}\tabularnewline
\toprule\noalign{}
Mobility Profile & Percentage \\
\midrule\noalign{}
\endfirsthead
\toprule\noalign{}
Mobility Profile & Percentage \\
\midrule\noalign{}
\endhead
\bottomrule\noalign{}
\endlastfoot
Local Resident & 77.1\% \\
Intra-City Commuter & 10.4\% \\
Inter-City Commuter & 12.1\% \\
Weekend-Only User & 0.4\% \\
\end{longtable}

\subsection{Naive vs.~Context-Aware
Comparison}\label{naive-vs.-context-aware-comparison}

The key comparison demonstrates the value of context-aware displacement
detection. We apply both methods to Aparri {[}population 68,839;
Philippine Statistics Authority (2021){]} across the 15-day
post-disaster period (Table~\ref{tbl-naive-comparison}).

Landfall occurred at approximately 5:00 PM on September 22, so the day
contains both pre-landfall and post-landfall hours. Sep 22's observed
values (the lowest weekday naive-CA gap at 1.6 pp against a weekday
average of 2.3 pp, and the lowest missing rate at 12.3\%) are consistent
with partial pre-landfall contamination, an incomplete day-1 response,
or both. As a sensitivity check, dropping Sep 22 raises the weekday mean
gap from 2.3 to 2.4 pp and turns ``all 11 of 11 weekdays show a gap''
into ``all 10 of 10''. The conclusion is robust to Sep 22's inclusion.
We retain it in the primary table as an indicative-not-clean boundary
day.

\begin{longtable}[]{@{}lllllll@{}}
\caption{Naive vs.~context-aware displacement rates for Aparri.
Difference in percentage points (pp). Missing = percentage of baseline
subscribers unobservable on that day. Upper = context-aware rate +
missing rate (worst-case scenario where all missing subscribers are
displaced).}\label{tbl-naive-comparison}\tabularnewline
\toprule\noalign{}
Date & Day & Naive & Context-Aware & Diff (N−CA) & Missing & Upper \\
\midrule\noalign{}
\endfirsthead
\toprule\noalign{}
Date & Day & Naive & Context-Aware & Diff (N−CA) & Missing & Upper \\
\midrule\noalign{}
\endhead
\bottomrule\noalign{}
\endlastfoot
Sep 22 & Mon & 8.3\% & 6.7\% & 1.6 pp & 12.3\% & 19.0\% \\
Sep 23 & Tue & 11.3\% & 9.2\% & 2.1 pp & 12.3\% & 21.5\% \\
Sep 24 & Wed & 9.0\% & 6.5\% & 2.5 pp & 14.2\% & 20.7\% \\
Sep 25 & Thu & 8.9\% & 6.3\% & 2.6 pp & 14.6\% & 20.9\% \\
Sep 26 & Fri & 8.8\% & 6.4\% & 2.4 pp & 14.8\% & 21.2\% \\
Sep 27 & Sat & 9.5\% & 9.4\% & 0.1 pp & 15.2\% & 24.6\% \\
Sep 28 & Sun & 9.3\% & 9.2\% & 0.1 pp & 21.8\% & 31.0\% \\
Sep 29 & Mon & 9.4\% & 6.9\% & 2.5 pp & 15.4\% & 22.3\% \\
Sep 30 & Tue & 9.7\% & 7.0\% & 2.7 pp & 15.8\% & 22.8\% \\
Oct 1 & Wed & 9.5\% & 6.9\% & 2.6 pp & 16.0\% & 22.9\% \\
Oct 2 & Thu & 9.4\% & 7.2\% & 2.2 pp & 21.0\% & 28.2\% \\
Oct 3 & Fri & 8.4\% & 6.4\% & 2.0 pp & 15.3\% & 21.7\% \\
Oct 4 & Sat & 9.6\% & 9.4\% & 0.2 pp & 16.0\% & 25.4\% \\
Oct 5 & Sun & 10.0\% & 9.9\% & 0.1 pp & 16.7\% & 26.6\% \\
Oct 6 & Mon & 9.8\% & 7.5\% & 2.3 pp & 16.4\% & 23.9\% \\
\end{longtable}

CV-based variability bounds of ±14\% apply to the context-aware rates
among observed users. The September 22 rate of 6.7\% has bounds of
5.8\%-7.6\%.\footnote{Aparri's baseline CV of 0.035 is the ratio of
  standard deviation to mean of the city-level daily subscriber count
  over May-August 2025, computed from a separate dataset supplied by
  Globe Telecom. Applying the disaster multiplier gives a disaster CV of
  0.035 × 2.0 = 0.071. Per Equation~\ref{eq-rate-ci}: 1.96 × 0.071 =
  0.139, giving bounds of ±14\% (i.e., 86\%-114\% of the estimate, the
  default row in Table~\ref{tbl-sensitivity}). For September 22: 6.7\% ×
  (1 ± 0.139) = 5.8\%--7.6\%.} Partial identification from unobserved
users is treated separately in §6.4.

The weekday-weekend contrast in Table~\ref{tbl-naive-comparison} (gap of
1.6-2.7 pp on weekdays, 0.1 pp on weekends) follows mechanically from
the rule design in Table~\ref{tbl-displacement-matrix}: the commuter
exception fires only for inter-city commuters on weekdays, so naive and
context-aware methods agree on weekends by construction. The contrast
confirms that the rule is implemented as intended and that inter-city
commuters are 12.1\% of the Aparri baseline
(Table~\ref{tbl-user-distribution}).

We applied the four uncertainty methods discussed in §5.7 to the Aparri
data. Average interval widths across the 15 post-disaster days are:
binomial (Clopper-Pearson, Wilson) 0.6 pp; CV-based 2.1 pp;
overdispersion-adjusted 3.8 pp; and Monte Carlo bootstrap 2.4 pp.~We
adopt the CV-based approach for the reasons given in §5.7.

\begin{figure}

\centering{

\includegraphics[width=1\linewidth,height=\textheight,keepaspectratio]{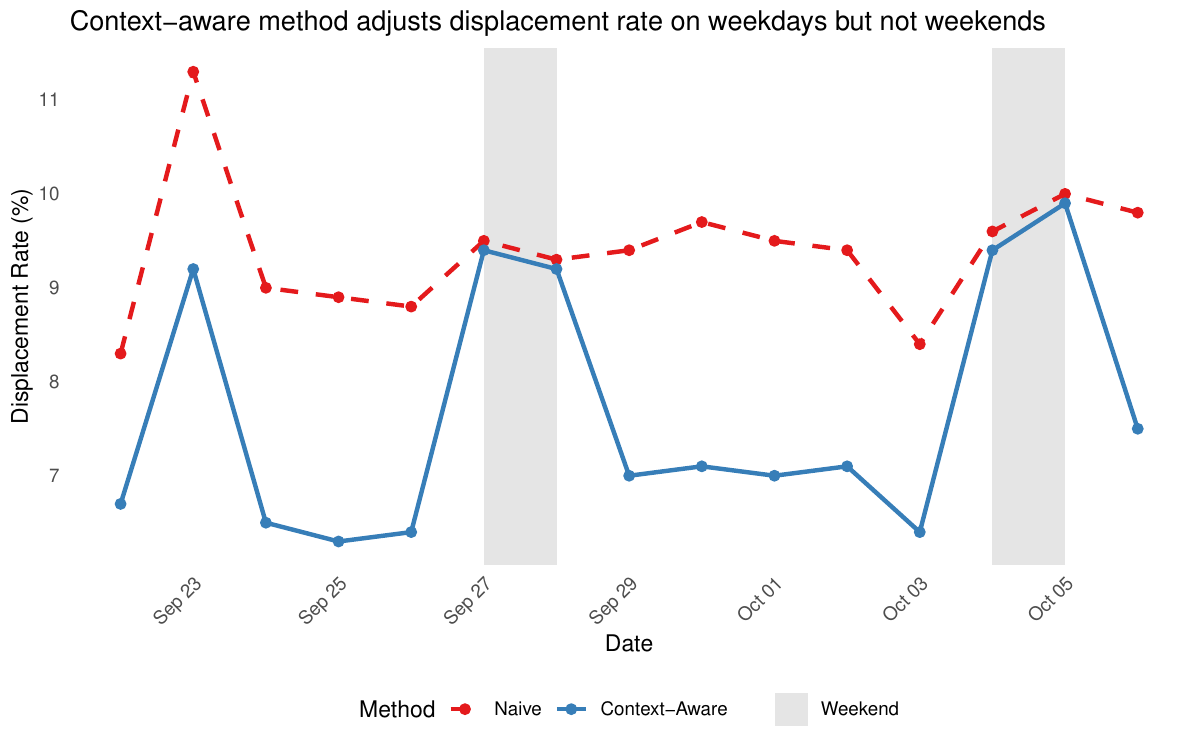}

}

\caption{\label{fig-timeseries}Naive vs.~context-aware displacement
rates for Aparri over the 15-day post-disaster period. On weekends
(shaded), the two methods converge. The weekday gap reflects commuter
misclassification by the naive method.}

\end{figure}

The weekday gap ranges from 1.6 to 2.7 pp (mean 2.3 pp,
Figure~\ref{fig-timeseries}). We report the pattern descriptively
because daily observations are temporally dependent. As a plausibility
check, we tracked inter-city commuters excused on Fridays: across the
two Friday-Saturday pairs in the observation period, 49-54\% returned to
their true home on Saturday, 37-42\% remained at the work city
(indeterminate from location data alone), and about 9\% moved to a third
location. The pattern is consistent with the commuter interpretation.
Formal validation against GPS ground truth (§7.2) is needed to attach
error rates with confidence.

\subsection{Scenario Bounds and Origin-Destination
Flows}\label{scenario-bounds-and-origin-destination-flows}

\begin{figure}

\centering{

\includegraphics[width=1\linewidth,height=\textheight,keepaspectratio]{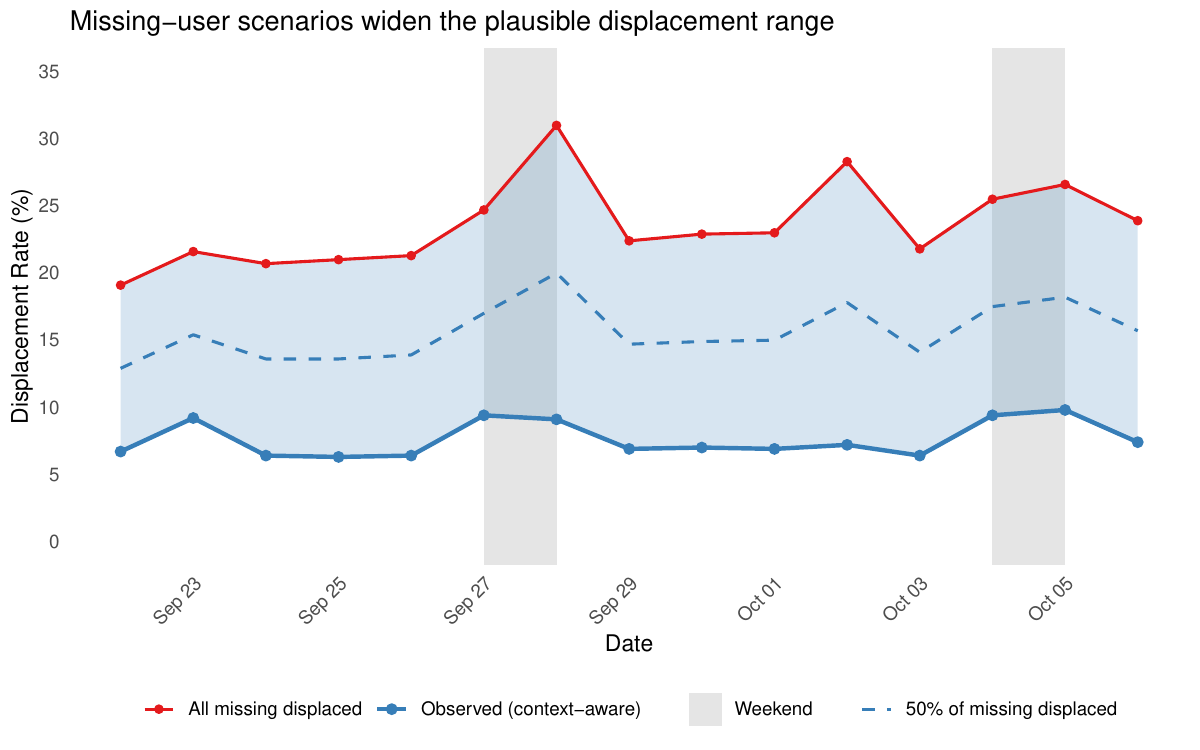}

}

\caption{\label{fig-scenario-bounds}Scenario bounds for Aparri
displacement rates over the 15-day post-disaster period. The observed
context-aware rate (blue solid line) represents the lower bound; the
upper bound (red line) assumes all missing subscribers were displaced.
The dashed line shows the 50\% scenario. Gray shading indicates
weekends.}

\end{figure}

Under the partial-identification framing of §5.7, the Upper column of
Table~\ref{tbl-naive-comparison} gives the daily worst-case scenario
(context-aware rate + missing rate). On September 22, the plausible
range is 6.7\%-19.0\%. By September 28, when the missing rate peaks at
21.8\%, the upper bound reaches 31.0\%. Figure~\ref{fig-scenario-bounds}
visualizes the envelope over the full 15-day period. Under a
middle-ground assumption (half of missing subscribers displaced),
displacement ranges from 12.9\% to 20.0\%.

Beyond displacement rates, the method produces origin-destination (OD)
flows showing where displaced persons went. We report September 23, the
first clean post-landfall weekday; we avoid September 22 because the
boundary-day caveats in §6.3 apply to its OD composition as well.

\begin{longtable}[]{@{}ll@{}}
\caption{Destinations for displaced persons from Aparri (September 23,
2025)}\label{tbl-destinations}\tabularnewline
\toprule\noalign{}
Destination & Share of Displaced \\
\midrule\noalign{}
\endfirsthead
\toprule\noalign{}
Destination & Share of Displaced \\
\midrule\noalign{}
\endhead
\bottomrule\noalign{}
\endlastfoot
Tuguegarao City & 23.9\% \\
Lallo & 21.8\% \\
Camalaniugan & 13.5\% \\
Buguey & 9.6\% \\
Santa Teresita & 6.0\% \\
Other (52 municipalities) & 25.1\% \\
\end{longtable}

The displacement pattern suggests two dynamics: displaced persons moved
to immediately neighboring municipalities (Lallo, Camalaniugan, Buguey,
Santa Teresita) as well as to Tuguegarao, the regional capital
approximately 80 km south. The movement to Tuguegarao may reflect access
to urban infrastructure and services, though this interpretation
requires further investigation.

\subsection{Comparison with Official
Estimates}\label{comparison-with-official-estimates}

The DSWD DROMIC Report \#14 (Department of Social Welfare and
Development, 2025), dated September 24, reported 10,663 displaced
persons from Aparri (15.5\% of the municipal population). Our
context-aware method estimates 6.5\% (5.6\%-7.4\%) between-municipality
displacement on the same date. The two figures are not directly
comparable: DSWD counts all evacuees from Aparri, including those who
sheltered within the municipality (at evacuation centers or with
relatives inside Aparri), while our method captures only subscribers
whose observed location shifted to a different municipality.
Within-municipality movement, which likely accounts for much of DSWD's
count, falls outside our detection scope.

The two approaches are complementary. Mobile data provides rapid
between-municipality origin-destination flows that field-based
enumeration cannot easily capture. DSWD provides ground-truth counts at
evacuation centers. Neither alone captures total displacement.

\subsection{Return Dynamics}\label{return-dynamics}

\begin{figure}

\centering{

\includegraphics[width=1\linewidth,height=\textheight,keepaspectratio]{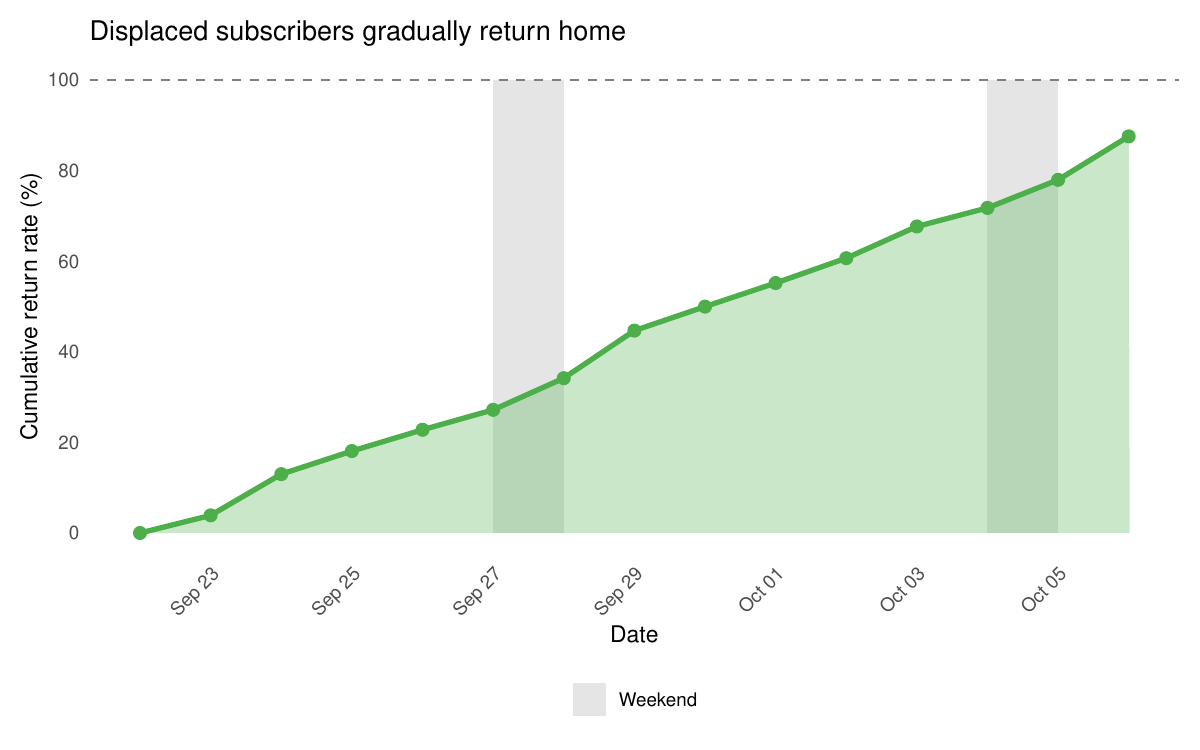}

}

\caption{\label{fig-return-dynamics}Cumulative return rate for Aparri
subscribers over the 15-day post-disaster period. The curve shows the
percentage of displaced subscribers who have returned to their expected
location, calculated per Equation~\ref{eq-cumulative-return}. Gray
shading indicates weekends.}

\end{figure}

Computed retrospectively at the end of the 15-day observation period
using Equation~\ref{eq-cumulative-return}, the cumulative return rate
trajectory rises monotonically from 0\% on September 22 to 87.6\% on
October 6 (Figure~\ref{fig-return-dynamics}, Table~\ref{tbl-returns}),
with 50\% reached by September 30 (day 9). The series is a post-hoc
recovery indicator rather than a real-time monitoring tool, because the
denominator \(\text{Max Displaced}_c\) depends on the full observation
period (§5.7). The interpretation is qualitative rather than a strict
unique-user fraction, since the underlying formula counts return events
and any re-displacement would inflate both numerator and denominator.
The remaining 12.4\% on October 6 reflects users not confirmed as
returned by that date, including those still displaced and those
unobservable.

\begin{longtable}[]{@{}lll@{}}
\caption{Cumulative return rate for Aparri subscribers, calculated per
Equation~\ref{eq-cumulative-return}.}\label{tbl-returns}\tabularnewline
\toprule\noalign{}
Date & Day & Cumulative Return Rate \\
\midrule\noalign{}
\endfirsthead
\toprule\noalign{}
Date & Day & Cumulative Return Rate \\
\midrule\noalign{}
\endhead
\bottomrule\noalign{}
\endlastfoot
Sep 22 & Mon & 0\% \\
Sep 23 & Tue & 3.9\% \\
Sep 24 & Wed & 13.0\% \\
Sep 25 & Thu & 18.1\% \\
Sep 26 & Fri & 22.8\% \\
Sep 27 & Sat & 27.2\% \\
Sep 28 & Sun & 34.2\% \\
Sep 29 & Mon & 44.7\% \\
Sep 30 & Tue & 50.0\% \\
Oct 01 & Wed & 55.2\% \\
Oct 02 & Thu & 60.7\% \\
Oct 03 & Fri & 67.7\% \\
Oct 04 & Sat & 71.8\% \\
Oct 05 & Sun & 78.0\% \\
Oct 06 & Mon & 87.6\% \\
\end{longtable}

\subsection{Sensitivity Analysis}\label{sensitivity-analysis}

We examine how displacement estimates respond to changes in key
parameters.

\begin{longtable}[]{@{}
  >{\raggedright\arraybackslash}p{(\linewidth - 8\tabcolsep) * \real{0.1974}}
  >{\raggedright\arraybackslash}p{(\linewidth - 8\tabcolsep) * \real{0.1711}}
  >{\raggedright\arraybackslash}p{(\linewidth - 8\tabcolsep) * \real{0.3158}}
  >{\raggedright\arraybackslash}p{(\linewidth - 8\tabcolsep) * \real{0.2105}}
  >{\raggedright\arraybackslash}p{(\linewidth - 8\tabcolsep) * \real{0.1053}}@{}}
\caption{Sensitivity of uncertainty bound width to disaster CV
multiplier, for the September 22 Aparri
estimate.}\label{tbl-sensitivity}\tabularnewline
\toprule\noalign{}
\begin{minipage}[b]{\linewidth}\raggedright
CV Multiplier
\end{minipage} & \begin{minipage}[b]{\linewidth}\raggedright
Disaster CV
\end{minipage} & \begin{minipage}[b]{\linewidth}\raggedright
Bounds (\% of estimate)
\end{minipage} & \begin{minipage}[b]{\linewidth}\raggedright
Relative Width
\end{minipage} & \begin{minipage}[b]{\linewidth}\raggedright
Change
\end{minipage} \\
\midrule\noalign{}
\endfirsthead
\toprule\noalign{}
\begin{minipage}[b]{\linewidth}\raggedright
CV Multiplier
\end{minipage} & \begin{minipage}[b]{\linewidth}\raggedright
Disaster CV
\end{minipage} & \begin{minipage}[b]{\linewidth}\raggedright
Bounds (\% of estimate)
\end{minipage} & \begin{minipage}[b]{\linewidth}\raggedright
Relative Width
\end{minipage} & \begin{minipage}[b]{\linewidth}\raggedright
Change
\end{minipage} \\
\midrule\noalign{}
\endhead
\bottomrule\noalign{}
\endlastfoot
1.5 & 0.053 & 90\%-110\% & ±10\% & -25\% \\
2.0 (default) & 0.071 & 86\%-114\% & ±14\% & --- \\
2.5 & 0.088 & 83\%-117\% & ±17\% & +24\% \\
\end{longtable}

The CV multiplier directly scales uncertainty bounds. Higher multipliers
produce wider bounds that better capture crisis-induced behavioral
variability but reduce estimate precision. Lower multipliers yield
tighter bounds but risk understating uncertainty during disasters. The
default value of 2.0 represents a pragmatic balance. Practitioners
should adjust based on disaster severity and data quality.

Other parameters affect displacement classification rather than
uncertainty quantification. We vary the two observation thresholds
across nine configurations: the weekend threshold is tested at 1-3 days
(out of \textasciitilde12 weekend days in the 40-day baseline), and the
weekday threshold at 3-7 days (out of \textasciitilde28 weekday days).
The ranges differ because fewer weekend days are available, so low
thresholds are already meaningful. Weekday thresholds below 3 would
produce unstable mode estimates. To match the main result (weekday
pattern across the 15-day post-disaster period), we report the
\emph{mean} across the 10 clean weekdays Sep 23 - Oct 6, excluding Sep
22 (§6.3). Across all nine configurations, the mean context-aware rate
ranges from 6.76\% to 7.22\% (a spread of 0.46 pp), and the mean
naive-context-aware gap stays within 2.24-2.38 pp.~The fundamental
pattern (weekday overestimation by the naive method) persists across all
configurations because it reflects the structural difference in handling
commuters rather than parameter-specific tuning.

\begin{longtable}[]{@{}
  >{\raggedright\arraybackslash}p{(\linewidth - 8\tabcolsep) * \real{0.1392}}
  >{\raggedright\arraybackslash}p{(\linewidth - 8\tabcolsep) * \real{0.1392}}
  >{\raggedright\arraybackslash}p{(\linewidth - 8\tabcolsep) * \real{0.3038}}
  >{\raggedright\arraybackslash}p{(\linewidth - 8\tabcolsep) * \real{0.1772}}
  >{\raggedright\arraybackslash}p{(\linewidth - 8\tabcolsep) * \real{0.2405}}@{}}
\caption{Sensitivity of the mean weekday naive-context-aware gap to
observation thresholds for Aparri, computed as the mean across 10 clean
weekdays Sep 23 - Oct 6. Bold row marks the default configuration. The
Baseline column shows relative change in qualifying
subscribers.}\label{tbl-threshold-sensitivity}\tabularnewline
\toprule\noalign{}
\begin{minipage}[b]{\linewidth}\raggedright
Weekend ≥
\end{minipage} & \begin{minipage}[b]{\linewidth}\raggedright
Weekday ≥
\end{minipage} & \begin{minipage}[b]{\linewidth}\raggedright
Baseline (vs default)
\end{minipage} & \begin{minipage}[b]{\linewidth}\raggedright
Mean CA Rate
\end{minipage} & \begin{minipage}[b]{\linewidth}\raggedright
Mean Gap (N − CA)
\end{minipage} \\
\midrule\noalign{}
\endfirsthead
\toprule\noalign{}
\begin{minipage}[b]{\linewidth}\raggedright
Weekend ≥
\end{minipage} & \begin{minipage}[b]{\linewidth}\raggedright
Weekday ≥
\end{minipage} & \begin{minipage}[b]{\linewidth}\raggedright
Baseline (vs default)
\end{minipage} & \begin{minipage}[b]{\linewidth}\raggedright
Mean CA Rate
\end{minipage} & \begin{minipage}[b]{\linewidth}\raggedright
Mean Gap (N − CA)
\end{minipage} \\
\midrule\noalign{}
\endhead
\bottomrule\noalign{}
\endlastfoot
1 & 3 & +4.1\% & 6.76\% & 2.29 pp \\
1 & 5 & +3.4\% & 6.83\% & 2.28 pp \\
1 & 7 & +3.2\% & 6.88\% & 2.24 pp \\
\textbf{2} & 3 & +1.6\% & 6.93\% & 2.33 pp \\
\textbf{2} & \textbf{5} & \textbf{default} & \textbf{7.07\%} &
\textbf{2.34 pp} \\
\textbf{2} & 7 & -0.6\% & 7.14\% & 2.31 pp \\
3 & 3 & +0.2\% & 6.98\% & 2.32 pp \\
3 & 5 & -2.4\% & 7.14\% & 2.36 pp \\
3 & 7 & -3.7\% & 7.22\% & 2.38 pp \\
\end{longtable}

The qualifying baseline varies by about 8 percentage points across
configurations, but the mean weekday gap remains stable at 2.24-2.38 pp.

The \textbf{nighttime window} (default 21:00-04:59) could not be tested
empirically because any nighttime/daytime weighting happens inside
Globe's vendor pipeline and is not a tunable parameter from our side.

\section{Quality Assurance and Future
Work}\label{quality-assurance-and-future-work}

\subsection{Internal Quality Controls}\label{internal-quality-controls}

The framework incorporates quality controls at each processing stage,
summarized in Table~\ref{tbl-quality-controls}. Observation thresholds
(≥2 weekend days, ≥5 weekday days) ensure stable mode-based location
assignments by filtering out users with insufficient data. The
residential baseline source flag (weekend versus weekday) tracks which
baseline informed each user's true-home assignment. Coverage monitoring
tracks user attrition from baseline to post-disaster periods, making
explicit how many users could not be classified.

\begin{longtable}[]{@{}
  >{\raggedright\arraybackslash}p{(\linewidth - 4\tabcolsep) * \real{0.3103}}
  >{\raggedright\arraybackslash}p{(\linewidth - 4\tabcolsep) * \real{0.3103}}
  >{\raggedright\arraybackslash}p{(\linewidth - 4\tabcolsep) * \real{0.3793}}@{}}
\caption{Internal quality controls embedded in the
framework}\label{tbl-quality-controls}\tabularnewline
\toprule\noalign{}
\begin{minipage}[b]{\linewidth}\raggedright
Control
\end{minipage} & \begin{minipage}[b]{\linewidth}\raggedright
Purpose
\end{minipage} & \begin{minipage}[b]{\linewidth}\raggedright
Reference
\end{minipage} \\
\midrule\noalign{}
\endfirsthead
\toprule\noalign{}
\begin{minipage}[b]{\linewidth}\raggedright
Control
\end{minipage} & \begin{minipage}[b]{\linewidth}\raggedright
Purpose
\end{minipage} & \begin{minipage}[b]{\linewidth}\raggedright
Reference
\end{minipage} \\
\midrule\noalign{}
\endhead
\bottomrule\noalign{}
\endlastfoot
Observation thresholds & Ensure stable mode-based location assignment &
Home Location Detection \\
Source flag (weekend/weekday) & Track residential baseline source &
Section~\ref{sec-residential-baseline} \\
Coverage monitoring & Track user attrition from baseline to
post-disaster & Equation~\ref{eq-missing-rate} \\
\end{longtable}

Practitioners should report three key metrics alongside displacement
estimates: the coverage rate (percentage of baseline users active on
each post-disaster day), the daily missing rate (percentage of baseline
users not observed on that specific day, per
Equation~\ref{eq-missing-rate}), and an attrition table showing user
counts at each processing step. The metrics allow consumers of the
estimates to understand data completeness and make informed judgments
about reliability.

\subsection{External Validation and Operational
Deployment}\label{external-validation-and-operational-deployment}

Two operational concerns require ongoing attention. First,
network-derived locations have not been validated against GPS ground
truth. A longitudinal volunteer study would compare GPS coordinates
against cell tower assignments and assess agreement between
self-reported commuter type and algorithm-assigned mobility profile.
Second, the relationship between mobile phone data and population
behavior drifts over time as cell tower networks change, mobile phone
penetration shifts, and commuting patterns evolve. Practitioners should
periodically recalibrate scaling factors, update
tower-to-administrative-unit mappings, and monitor baseline CV trends to
detect drift. Detailed validation protocols and drift-monitoring
guidelines are operational documentation outside the scope of this
paper.

\section{Discussion}\label{discussion}

\subsection{Strengths}\label{strengths}

The framework's primary contribution is explicit handling of commuter
populations through mobility profile classification and context-aware
detection, designed to reduce false positives from regular commuting.
Three complementary metrics (rates, origin-destination flows, return
dynamics) support both immediate response and recovery tracking, and an
aggregation-based approach preserves individual privacy by keeping
trajectories out of outputs. Operational uncertainty bounds derived from
baseline CV offer decision support rather than formal statistical
inference (§5.7). The false negative problem noted in §2.5 remains
inherent to all location-based approaches.

\subsection{Case Study Implications}\label{case-study-implications}

The Aparri case study reveals three operational insights.

First, the naive method's weekday overcount of 1.6-2.7 percentage
points, if the commuter exception is correct, represents daily
misclassification that could misallocate evacuation resources for
responders relying on uncorrected estimates.

Second, mobile data provides origin-destination flows that evacuation
center registration does not systematically capture. Responders can
identify where displaced persons went, enabling targeted resource
allocation to receiving communities.

Third, persistent displacement over 15 days (rising from 6.7\% to 7.5\%
when comparing the same weekday) signals that emergency response must
extend beyond immediate relief to support prolonged displacement and
recovery.

\subsection{Limitations}\label{limitations}

Several limitations warrant acknowledgment. Mobile phone data introduces
coverage biases: ownership skews toward younger, urban, and
higher-income populations, so estimates may underrepresent elderly,
rural, and low-income groups who can be particularly vulnerable in
disaster contexts. Aggregate mobility estimates have been shown to be
robust to ownership biases in some settings (Wesolowski et al., 2013),
though the robustness varies by context.

Additionally, single-operator data (as used in the case study)
introduces bias, as market share varies geographically and
demographically. Multi-operator data would improve representativeness
but is rarely available. We report displacement as percentages rather
than absolute subscriber counts to protect proprietary data. The spatial
resolution of cell tower data is also coarser than GPS, limiting
precision in assigning users to administrative units, especially in
rural areas with sparse tower infrastructure.

The nighttime-as-home heuristic assumes people sleep at their residence,
which may not apply to night-shift workers or those in informal housing
arrangements. Similarly, the weekend-as-residence assumption may be less
reliable in contexts where weekend work is common or cultural practices
differ. Multi-SIM usage poses an additional challenge: individuals with
multiple SIM cards may appear as separate users, potentially inflating
baseline counts or creating inconsistent home assignments across SIMs.

The framework operates at daily resolution by design, defining
displacement based on where a user is observed on a given day relative
to their expected location. The approach prioritizes robustness over
granularity. Intra-day tracking would require denser observations and
introduce additional noise. Practitioners should note that if a
post-disaster day coincides with a national holiday, results for that
day may warrant a caveat or exclusion from trend analysis, as holiday
mobility patterns often resemble weekends regardless of the calendar
day. (The baseline period already excludes major holidays and recent
disasters, as described in the Temporal Framework section.)

The disaster CV multiplier of 2.0 represents a provisional estimate
intended to account for increased behavioral variability during crises.
Empirical calibration against observed displacement events would
strengthen the parameter choice. Users who cannot be observed in the
post-disaster period pose an inherent limitation: they cannot be
classified as displaced or not displaced, so estimates are necessarily
conditional on continued network activity. Finally, the framework has
been developed and illustrated with a single disaster type; application
to other contexts such as conflict or slow-onset disasters requires
validation.

\subsection{Generalizability}\label{generalizability}

The framework is designed for sudden-onset disasters with a defined
onset date, but can be adapted to other contexts with modification.
Slow-onset disasters such as droughts would require a rolling baseline
approach and gradual displacement detection rather than a discrete
pre/post comparison. Conflict settings may necessitate security-aware
data collection protocols and shorter baseline periods due to rapidly
changing conditions. Different geographic contexts may require
recalibration of spatial resolution expectations based on local tower
density, while different mobile markets warrant adjustment of scaling
factors and assessment of coverage biases specific to that population.

\section{Conclusion}\label{conclusion}

We present a methodological framework for estimating population
displacement from mobile phone data that addresses the systematic bias
introduced by ignoring regular commuting patterns. Through mobility
profile classification and context-aware displacement detection, the
framework is designed to reduce commuter misclassification. Independent
validation of the rule's accuracy remains future work.

The framework produces three complementary metrics: displacement rates,
origin-destination flows, and return dynamics, scaled to population
level with operational uncertainty bounds derived from baseline
variability. Built-in quality assurance protocols and transparent
parameter specification support reproducible application across
contexts.

Key contributions include:

\begin{enumerate}
\def\labelenumi{\arabic{enumi}.}
\tightlist
\item
  Explicit handling of commuter populations through mobility profile
  classification
\item
  Day-of-week sensitive displacement detection rules
\item
  CV-based uncertainty quantification with disaster adjustment
\end{enumerate}

Future work includes empirical calibration of the disaster CV
multiplier, validation of the commuter rule against GPS ground truth,
implementation of the internal-detection mode (§5.4), application to
additional disaster types to assess generalizability, and development of
bounded uncertainty formulations such as a Beta-based approach. An
additional extension could incorporate a tunable parameter representing
the proportion of missing users assumed to be displaced, allowing
practitioners to adjust estimates based on context.

\section{Author Contributions
(CRediT)}\label{author-contributions-credit}

\textbf{Rajius Idzalika}: Conceptualization, Methodology, Software
(original research codes), Validation (production pipeline
verification), Writing -- original draft, Supervision, Project
administration.

\textbf{Muhammad Rheza Muztahid}: Conceptualization, Methodology,
Software (production pipeline), Validation (paper-to-code verification).

\textbf{Radityo Eko Prasojo}: Supervision, Writing -- review \& editing.

\section{Acknowledgments}\label{acknowledgments}

Work described here was conducted as part of the Data Insights for
Social and Humanitarian Action (DISHA) initiative led by UN Global Pulse
and made possible by financial support from Google.org and the Patrick
J. McGovern Foundation.

This work builds on the foundation established by Pulse Lab Jakarta's
pioneering research on mobile phone data for development and
humanitarian applications, which developed methodologies and
partnerships that enabled this framework.

We gratefully acknowledge our partnership with Globe Telecom, which
provided access to de-identified mobile phone data and enabled empirical
application and iterative refinement of the framework.

We thank McKinsey for their collaboration on earlier displacement
estimation work using an older version of the mobile data, which
informed the development of this framework.

\section{References}\label{references}

\phantomsection\label{refs}

\end{document}